\documentclass[prd, twocolumn, showpacs, superscriptaddress, floatfix, nofootinbib]{revtex4}

\usepackage[pdftex]{graphicx}

\usepackage{epsf}
\usepackage{amsmath}
\usepackage{amssymb}

\usepackage{graphicx}
\usepackage{dcolumn}
\usepackage{bm}
\def\be{\begin{equation}}
\def\ee{\end{equation}}
\def\bea{\begin{eqnarray}}
\def\eea{\end{eqnarray}}

\begin{document}





\title{Searching for Cosmic Strings in New Observational Windows}


\author{Robert H. Brandenberger}
\email{rhb@physics.mcgill.ca}
\affiliation{Physics Department, McGill University, 3600 University Street, Montreal, QC, H3A 2T8, Canada}


\begin{abstract}

Cosmic strings are predicted in many models beyond the Standard Model of particle
physics. In models which admit strings, a network of strings will inevitably be formed in a phase
transition in the early universe and will persist to the present time. Strings leave
behind distinctive features in cosmology. Searching for these signatures in new
observational windows provides a way to constrain particle physics at the high energy
scale and is thus complementary to searches for new physics at the low energy end,
for example at the LHC. Specifically, I will discuss signatures of cosmic strings in
cosmic microwave background polarization maps and in 21cm redshift surveys.

\end{abstract}

\maketitle



\newcommand{\eq}[2]{\begin{equation}\label{#1}{#2}\end{equation}}

\section{Introduction}
\label{Introduction}

Cosmic strings \cite{Kibble1} are linear topological defects which arise 
in a range of relativistic quantum field theories (for reviews see e.g 
\cite{CSrevs}). Good analogs of cosmic strings are
vortex lines in superfluids and superconductors. Line defects
in crystals can be viewed as another analog system. Cosmic strings
form lines of trapped energy density, and this energy density can
curve space-time and have important effects in cosmology \cite{ZelVil}.

Cosmic strings are predicted to form in many particle physics models
beyond the Standard Model. In particular, they are predicted to form
at the end of inflation in many inflationary models, e.g. supergravity
models \cite{Rachel} and brane inflation models \cite{braneinflation}.
Cosmic strings may also survive as cosmic superstrings in alternatives
to inflation such as ``String Gas Cosmology" \cite{SGC}. The key point
for cosmology is that in any field theory model which admits cosmic
string solutions, a network of strings inevitably forms at some point
during the early universe \cite{Kibble2}, and it persists to the present time. Hence,
the detection of cosmic strings would give us information about
particle physics at very high energy scales.

Since cosmic strings are relativistic objects, a straight string is described 
by one number, namely its mass per unit length $\mu$ which also equals its
tension, or equivalently by the dimensionless number $G \mu$, where
$G$ is Newton's gravitational constant (we are using units in which
the speed of light is $c = 1$). In simple quantum field theory models
the tension is related to the energy scale $\eta$ at which the strings
are formed via $\mu \sim \eta^2$ (see the following section for a more
precise discussion). The cosmological signatures of strings
are thus more substantial for larger values of $\mu$ which implies
larger values of the energy scale $\eta$. Hence, searching for cosmological
signatures of strings is a tool to probe particle physics beyond the
Standard Model at the highest energy scales (as opposed to accelerator
experiments like the LHC which probe new physics at low energy
scales). 

In fact, current limits on cosmic strings already \cite{Dvorkin} provide a
constraint
\be
G \mu \, < \, 1.5 \times 10^{-7}
\ee
which  rules out certain Grand Unified particle physics models with
very high scale symmetry breaking. This limit comes from the observational
upper bound on the contribution of cosmic strings to the angular power
spectrum of cosmic microwave background (CMB) anisotropies obtained
by combining the results of the WMAP satellite \cite{WMAP} with those of the South
Pole Telescope \cite{SPT} (see also \cite{Bevis} for a comparable limit obtained
by combining results from WMAP and from the Atacama Cosmology Telescope
\cite{ACT}, and \cite{other} for earlier limits).

Given the constraints on particle physics models which can be derived from
current observations, it is of great interest to try to improve the observational 
upper bounds on the cosmic string tension since this will allow us to constrain
high energy scale particle physics models more strongly than what is possible today.

Cosmic strings can also produce many good things for cosmology in addition
to contributing to cosmological structure formation. Cosmic strings may play
a role in baryogenesis (see e.g. \cite{CSBG}). Certain types of strings can 
provide a mechanism for the production \cite{CSB} of seed magnetic fields which are
coherent on galactic scales \footnote{The challenge for particle physics models
of magnetogenesis is to obtain the observed coherence length since microphysics
typically produces coherence lengths which are much too small. In the case of
cosmic string seeds, the increase of the comoving curvature radius of the long
strings provides the mechanism of obtaining a large magnetic field coherence
length.} Cusps
on cosmic string loops can also yield a contribution to ultra-high-energy cosmic
rays \cite{Jane, Yifu}. Finally, cosmic string loops may assist in the
assembly of the large mass concentrations required to seed super-massive
black holes.

For all of the above reasons it would thus be wonderful to have evidence for
the existence of cosmic strings in nature. The search for cosmic strings is
therefore of great interest independent of whether the search in fact
finds signals of cosmic strings. If it does, then we will have discovered
something completely new in the universe. If it does not, then we will have
derived tighter constraints on particle physics at very high energy scales.

In this talk I will review recent work on signatures of cosmic strings in new
observational windows. Up to the present time, the tightest and most
robust constraints on the cosmic string tension have come from analyses
of CMB temperature maps. Here, I will focus on the signatures of strings
in CMB polarization maps and 21cm redshift surveys, two emerging windows
to explore the cosmos. 

The main points to take away from this talk are the following. Firstly,
cosmic strings lead to nonlinearities already at very high redshifts. Hence,
the signatures of cosmic strings are more pronounced at higher than at
lower redshifts where they are masked by the nonlinearities produced 
by the Gaussian density fluctuations which must be present and which
dominate the total power spectrum of cosmological perturbations.
Secondly, cosmic strings lead to perturbations which are highly 
non-Gaussian and which predict specific geometrical patterns in
position space. By computing a power spectrum, information about
these patterns are lost. Hence, tighter limits on the cosmic string
tension can be obtained if we analyze the data in position space.
Thirdly, 21cm redshift surveys appear to be an ideal window to search
for cosmic string signatures \cite{Oscar1}.

The outline of this talk is as follows. We first present a brief review of
the basics of cosmic strings. In Section 3 we introduce the two main
mechanisms which will play a role in determining the cosmic string
signals in observations, namely the Kaiser-Stebbins \cite{KS} (see
also \cite{Gott}) lensing effect and the cosmic string wake \cite{wake}.
We also briefly review the well-known resulting signal of strings in
CMB temperature maps. The key sections of this talk are Sections 4 and
5. In the first, we discuss the signal of a long straight cosmic string
in CMB polarization maps, and in the second we turn to the signal in
21cm redshift surveys. 

\section{Cosmic String Review}
\label{Review}

In a class of relativistic quantum field theories, cosmic strings form 
after a phase transition in the early universe during which an internal
symmetry in field space is spontaneously broken. Let us consider a
simple toy model involving a complex scalar field $\phi$ with
potential
\be
V(\phi) \, = \, \frac{\lambda}{4} \bigl( |\phi|^2 - \eta^2 \bigr)^2 \, 
\ee
where $\eta$ is the vacuum expectation value of the modulus of
$\phi$ and $\lambda$ is a coupling constant.

To determine whether a particular field theory admits cosmic string 
solutions or not, the key concept is that of the {\it{vacuum manifold}}
${\cal{M}}$, the set of field values which minimizes the potential. In the
above example ${\cal{M}}$ is homotopically equivalent to the circle $S^1$.

In thermal equilibrium, the potential obtains finite temperature corrections
\cite{Dolan, Linde} (see e.g. \cite{RMPrev} for a review). Specifically, there
is an extra contribution
\be
\Delta V_T(\phi) \, \sim \, T^2 |\phi|^2 \, ,
\ee
where $T$ is the temperature. Hence, there is a critical temperature $T_c$
above which the lowest potential energy state is $\phi = 0$ and the
field symmetry (rotation in the complex field plane) is unbroken. Thus, in
the early universe the average value of $\phi$ at each point in space will
be $\phi = 0$, but as the universe cools below the temperature $T_c$
this state becomes unstable and at each point in space $\phi$ will
want to roll down the potential to take on a value in ${\cal{M}}$.

The key point is \cite{Kibble2} that by causality there can be
no correlation between the field values in ${\cal{M}}$ which are taken
on at points in space which are out of causal contact, i.e. which
are further apart than the Hubble distance $t$. Hence, there is
a probability of order $1$ that the field values in ${\cal{M}}$
evaluated for a loop ${\cal{C}}$ in space of radius $t$ will form an incontractible
loop in ${\cal{M}}$. This implies that there must be a point in
space on any disk bounded by ${\cal{C}}$ where $\phi = 0$.
Around these points there is trapped potential and spatial
gradient energy. 

Energy minimization arguments make it obvious
that the distinguished points on different disks 
(with the same boundary) form part of a line.
This line is the cosmic string. It is a line of points with $\phi = 0$
surrounded by a tube of trapped energy. The width $w$ of the tube
is determined by minimizing the sum of potential and spatial
gradient energy, leading to the result
\be
w = \lambda^{-1/2} \eta^{-1}
\ee
such that the resulting mass per unit length $\mu$ is independent of
the coupling constant:
\be
\mu = \eta^2
\ee
(up to numerical factors independent of $\lambda$).

To recap, causality tells us that at the symmetry breaking phase transition
at time $t = t_c$ a network of cosmic strings will form in any field theory model which
admits cosmic string solutions. Cosmic strings are closed - they
cannot have ends. Hence, they are either string loops or infinite in
length. Note that causality in fact ensures that (in an infinite space) 
strings of infinite length will be formed. A way to picture the network
of strings is as a random walk with typical curvature radius $\xi(t)$ which
is bounded from above by the Hubble length $t$.

The same causality argument which is used to predict the formation
of strings can be used to show that the network of strings will survive
at all times $t > t_c$, including the present time. Hence, cosmic strings
formed in the very early universe will lead to signatures which can be
searched for in current cosmological observations.

We will work in terms of a one-scale model \cite{CSearly}
of the cosmic string distribution
which is given by the correlation length $\xi(t)$ which describes both the
mean curvature radius and the separation of the ``long" strings (``long"
meaning with length greater than $t$). Kibble's causality argument
tells us that $\xi(t) < t$. Dynamical arguments show that $\xi(t)$ cannot
be much smaller than $t$. The argument is as follows: cosmic strings are
relativistic objects. If the curvature radius is smaller than $t$, then the
strings will move with relativistic speed. Hence, intersections of strings
will occur. It can be shown \cite{Paul} that unless the relative speed of 
two intersecting string segments is extremely close to the speed of light, 
then the string segments will not cross, but they will intersect and 
exchange ends. In this way, the long strings can form string loops, 
which in turn oscillate and decay by emitting gravitational radiation. 
This process is described by a Boltzmann equation (see e.g. \cite{CSrevs})
from which it follows that if $\xi(t) \ll t$ then
$\xi(t) / t$ will increase, and that there will be a dynamical fixed point
{\it scaling solution} with
\be
\xi(t) \, \sim t \, .
\ee
The approach to a scaling solution has been confirmed in a series
of numerical simulations \cite{CSsimuls}. The process of long string
intercommutation leaves behind a distribution of cosmic string
loops which also achieves a scale-invariant form. We will
not discuss this issue since we will focus on the cosmological
signatures of the long strings.

Since cosmic strings carry energy, they can gravitate and produce
cosmological signals. This was first pointed out in \cite{ZelVil}.
In the early 1980s cosmic strings were studied as an alternative
to cosmological inflation \cite{CSearly}. Cosmic string formation
produces entropy fluctuations on super-Hubble scales which then
seed a curvature fluctuation which grows on super-Hubble scales.
Such fluctuations are called ``active". They are also ``incoherent"
meaning that there are no phase correlations in phase space between
different Fourier modes \cite{active}.
Hence, although cosmic strings predict a spectrum of density
fluctuations which is scale-invariant and hence an angular power
spectrum of CMB anisotropies which is also scale-invariant on large
angular scales \cite{CSspectrum}, the acoustic oscillations which 
are characteristic of models like inflation where a scale-invariant 
spectrum of adiabatic
fluctuations are generated whose amplitude remains constant on
super-Hubble scales are absent \cite{active}. Once the acoustic
oscillations were discovered by various experiments such as the
Boomerang experiment \cite{Boomerang}, interest in cosmic
strings collapsed.
 
However, given the realization that cosmic strings are produced after
inflation in many particle physics models, interest in searching
for observational signatures of cosmic strings as a supplementary
(not main) source of structure in the universe has increased. The
search for strings in the cosmos is emerging as a promising way
to probe physics at energy scales much larger than those that
can ever be reached in terrestrial accelerators, as already mentioned
above. Independent of whether strings are in fact discovered or
the bounds on the existence of strings are improved we will have
learned a lot about particle physics.
 
\section{Kaiser-Stebbins Effect and Cosmic String Wakes}
\label{Wakes}

In this talk I will focus on gravitational effects of cosmic strings. These
are based on two main ``actors", firstly the {\it Kaiser-Stebbins} lensing
effect \cite{KS, Gott}, and secondly the existence of {\it string wakes}
\cite{wake}. 

The string lensing effect is based on the fact that a long straight
cosmic string with equal tension and energy per unit length
leads to a conical structure of space perpendicular to the
string \cite{deficit}. Unwrapping the cone onto a plane leads
to a ``deficit angle"
\be
\alpha \, = \, 8 \pi G \mu \, .
\ee
If we now consider a cosmic string with transverse velocity $v$
and associated relativistic gamma factor $\gamma(v)$, then
if we look at the CMB in direction of the string (see
Fig. 1), then photons
passing on the two sides of the string are observed with a
relative Doppler shift
\be
{{\delta T} \over T} \, = \, 8 \pi \gamma(v) v G \mu \, .
\ee
This leads to a line discontinuity in the CMB sky, the linear
direction corresponding to the projection of the tangent
vector of the string onto our last light cone \cite{KS, Gott}.

\begin{figure}
\includegraphics[height=8cm]{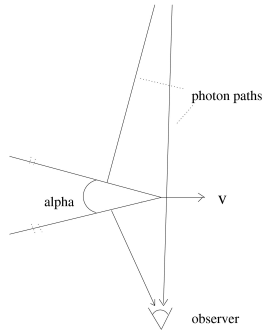}
\caption{Sketch of the geometry of space perpendicular to a
long straight string segment. Space is conical which corresponds
to a deficit angle in the plane. If the string is moving with
velocity $v$ in a direction transverse to the direction of the
string, then photons passing on different sides of the string
are measured with a relative Doppler shift. This is the
Kaiser-Stebbins effect.}
\label{fig1}
\end{figure}

Given a scaling distribution of strings, each string which
intersects our past light cone will lead to a line discontinuity
in the CMB sky. As shown in \cite{Joao}, for strings formed
at a phase transition at time $t_c$ in the early universe (as opposed
to idealized strings being present at all times), the deficit
angle at time $t$ has finite depth $d(t)$, namely
\be
d(t) \, \simeq \, (t - t_c) \, \simeq \, t
\ee
for $t \gg t_c$. Hence, the signature of an individual
string in the sky corresponds to a rectangle in the sky
with the above value of $\delta T$. The
dimensions of the rectangle are the angles
given by the comoving distances corresponding to
length $c_1 t$ and depth $t$, where $c_1$ is
a constant of order $1$. 

We will be making use of a toy model for the (long) cosmic
string scaling first introduced in \cite{Periv} and widely
used since then in cosmic string research. We replace
the infinite string network at time $t$ by a set of straight string 
segments of length $c_1 t$ whose center of mass positions,
tangent vectors and velocity vectors are randomly
distributed and uncorrelated. These string segments live
for one Hubble expansion time (the average time between
string intercommutations). In the subsequent Hubble time
steps there are new string segments with larger length
whose centers, tangent and velocity vectors are assumed
to be uncorrelated with those of the string segments in the
previous Hubble time step. As a reflection of the scaling of
the string network, we consider a fixed number $N$ per
string segments in each Hubble volume. The number $N$
can be determined by comparing with numerical simulations.
Causality tells us that $N \geq 1$. Numerical simulations
give values of the order $N \sim 10$ \cite{CSsimuls}.

String segments in each Hubble time step contribute to the
total Kaiser-Stebbins effect. String segments in the first
Hubble time step after recombination are the most numerous.
The size in the sky of the corresponding temperature patches
is about one degree. Strings from later Hubble time steps
are less numerous but larger (see Fig. 2).

\begin{figure}
\includegraphics[height=14cm]{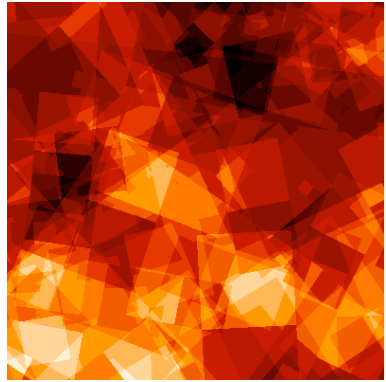}
\caption{CMB anisotropy map for a $10^{o} \times 10^{o}$ 
patch of the sky at $1.5^{'}$ resolution (the specifications 
are chosen to be comparable to those of the SPT and ACT 
telescopes - in fact both of these telescopes map a larger 
fraction of the sky) in a model in which the fluctuations are 
given by a scaling distribution of cosmic strings. The
color coding indicates the amplitude of the temperature
anisotropy.} 
\label{fig2}
\end{figure}

To identify the signal of cosmic strings in CMB temperature
maps, good angular resolution is more important than full sky
coverage. Hence, experiments such as SPT and ACT will
yield better limits than even the Planck satellite. 

The distinctive edges in CMB temperature maps produced
by strings are washed out in the angular power spectrum. Hence,
tighter limits can be set on the string tension by analyzing the
CMB maps in position space using edge detection algorithms
rather than by simply computing a power spectrum. Specifically,
by using a numerical implementation of the Canny edge detection
algorithm \cite{Canny} it appears that a bound one order of
magnitude stronger than the current bound might be achievable
by using data from the SPT telescope \cite{Rebecca} (see also
\cite{Berger, Stewart} for initial work on applying the Canny
algorithm to CMB maps). The projected bound is
\be
G \mu \, \le \, 2 \times 10^{-8} \, .
\ee

The second main ``actor" in the story presented here is the cosmic
string wake \cite{wake}, Consider a long straight string segment
moving through the uniform matter distribution of the early universe.
From the point of view of a point behind the moving string, it appears
as though matter acquires a velocity perturbation
\be
\delta v \, = \, 4 \pi v \gamma(v)  G \mu
\ee
from above and below towards the plane behind the moving string. This
in turn leads to a wedge-shaped overdensity (twice the background
density) behind the string. This is the {\it wake}.

Working again in the context of the toy model of \cite{Periv}, each string
segment in each Hubble expansion time generated a wake. Consider
a string at time $t_i$. The physical dimension of the induced wake at the
time $t_i$ will be
\be
c_1 t_i \times v \gamma(v) t_i \times 4 \pi  v \gamma(v) G \mu t_i \, ,
\ee
where the first factor is the size along the tangent vector of the string,
the second factor is the depth (in direction opposite to the string motion),
and the third factor is the mean width. At the leading edge (the instantaneous
location of the string), the width of the wake is zero, whereas at the trailing
edge (the initial position of the string segment) the width  is twice the mean
width $w$.

\begin{figure}
\includegraphics[height=4.5cm]{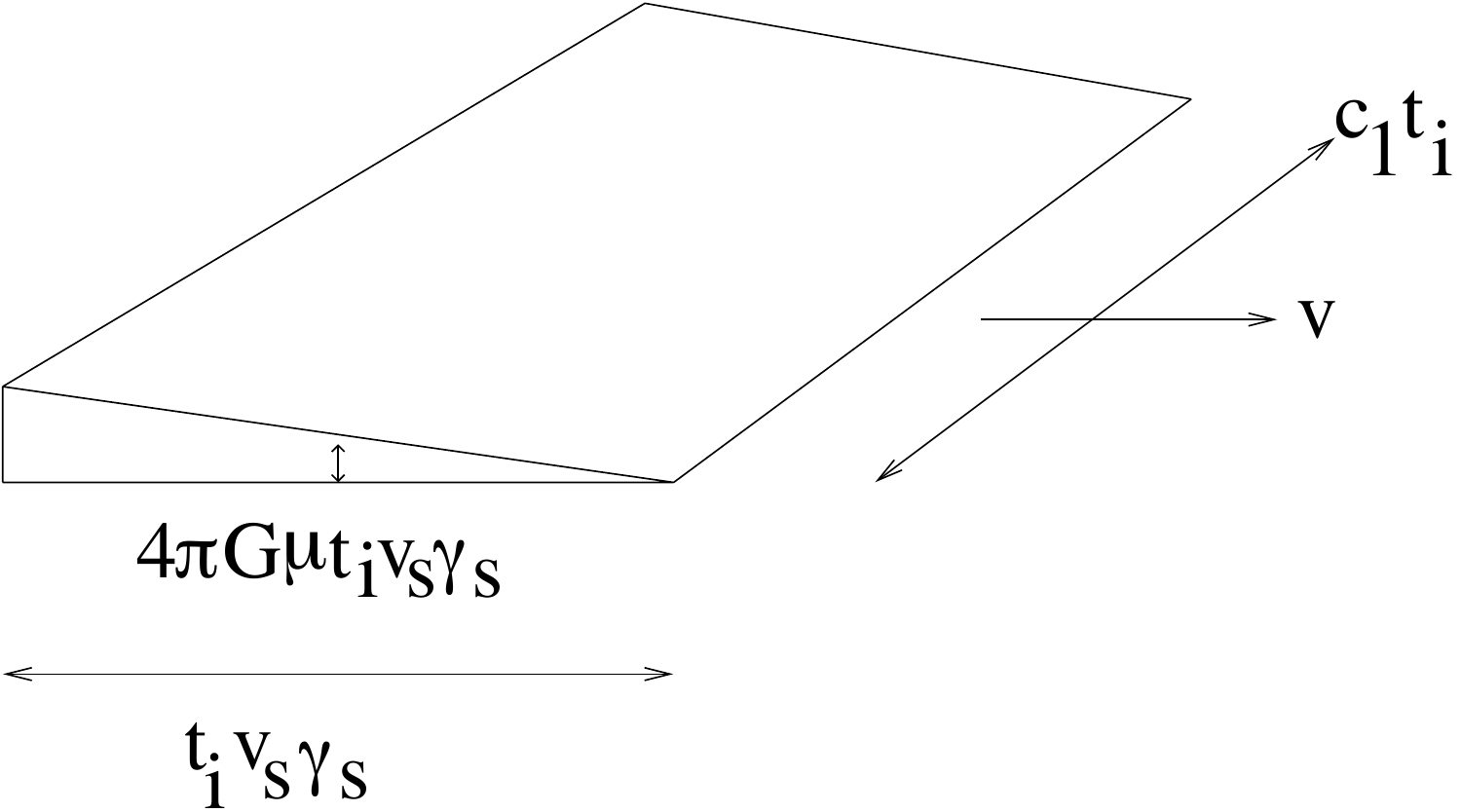}
\caption{Geometry of a cosmic string wake. Such a wake is extended
in the plane spanned by the direction tangential to the string segment 
and by the velocity vector, while its initial size perpendicular to this
plane is thin. In the figure, $v_s$ is the string velocity $v$,
and $\gamma_s$ is its related gamma factor.} 
\label{fig3}
\end{figure}

Once formed, a wake will grow in thickness via gravitational accretion. This
process can be studied using the Zel'dovich approximation \cite{Zel}. The
idea is to consider a shell of matter which is located initially (at the time 
$t_i$ when the wake is laid down) at a physical height
$h(t_i) = a(t_i) q$ above the center of the wake. The quantity $q$ is the
initial comoving height. As a consequence of the gravitational pull of
the matter overdensity in the wake, a comoving displacement $\psi(t)$
gradually builds up (where $\psi(t_i) = 0$). The physical height at
time $t > t_i$ then can be written as
\be 
h(q, t) \, = \, a(t) \bigl( q - \psi \bigr) \, , \nonumber
\ee
where $a(t)$ is the cosmological scale factor.
The time evolution of the height is then determined via Newtonian gravity
\be
{\ddot h} \, = \, - \frac{\partial \Phi}{\partial h} \, ,
\ee
where $\Phi$ is the Newtonian gravitational potential which is determined
via the Poisson equation in terms of the mass overdensity. We then
calculate the value $q(t)$ (which we call $q_{nl}(t, t_i)$) 
for which the shell is ``turning around" at time $t$, i.e.
\be
{\dot h(q(t), t)} \, = \, 0 \, .
\ee
After turnaround, the shell will virialize at a physical height which is half
the turnaround height \footnote{Note that this picture of cosmic string wake
growth has been confirmed by an Eulerian hydro simulation \cite{Sorn}.} 
This virialized region forms the wake. The result
of a straightforward calculation shows that (in agreement with what
follows from linear perturbation theory), the comoving height of the wake
grows linearly in the scale factor, i.e.
\be
q_{nl}(t, t_i) \, = \, \frac{a(t)}{a(t_i)} \frac{24 \pi}{5} v \gamma(v) G \mu (z(t_i) + 1)^{-1/2} t_0 \, ,
\ee
where the expression on the right hand side is the ratio of scale factors multiplying the
initial comoving width of the wake (modulo a factor of order $1$). Note
that $z(t)$ is the cosmological redshift.  
In the context of cosmic strings this analysis was originally done in
\cite{CSZel1} (accretion onto loops) and \cite{CSZel2} (accretion onto
wakes).

Since the turnaround height itself is half the height
the shell would have if it were simply to expand with the Hubble flow,
the resulting overdensity in the wake is a factor of $4$. Note that
for accretion onto a string loop, the resulting overdensity is $64$
since there is contraction in all three spatial directions (see e.g.
\cite{Pagano}).

Let us end this section with a couple of comments. First of all,
the planar dimensions of the wake will retain constant comoving
size. Secondly, there is an important difference between the lensing
signal due to string segments and the wake signal. Since the
string segment only lives for one Hubble expansion time, only
strings whose finite time world sheet intersects the past light cone
lead to an observable signal. On the other hand, wakes persist
even after the string segment which has seeded them has decayed.
Hence, all string segments within the past light cone lead to
observable wake signals. In the following, it is signals due to
wakes which will be discussed.

\section{Signatures of Cosmic Strings in CMB Polarization}
\label{Polarization}

When primordial CMB quadrupole radiation scatters off of a
gas cloud, the residual free electrons in the cloud lead to
polarization.  Wakes are regions of enhanced density, and hence also of 
enhanced free electron density. Photons emitted at the time of 
recombination acquire extra polarization when they pass through
a string wake. It is this signal which we study here \cite{Holder}.

There are two polarization modes - E and B modes.
A Gaussian random field of density fluctuations leads to pure E-mode
polarization. In contrast, cosmic string wakes lead to a statistically
equal distribution of E and B-mode polarization. The reason for
the generation of a B-mode component is that there is a
distinguished vector given by the normal vector to the wake.
This vector is independent of the CMB quadrupole vector.

The amplitude $P$ of the polarization signal depends on the
column density of free electrons which the CMB photons
encounter when they cross the wake. If $t$ is the time when
the photons are passing through the wake, the column
density is proportional to the residual ionization fraction $f(t)$ 
at time $t$, the number density of baryons and the width of
the wake. The value of $P$ is then determined by multiplying
the result with the scattering cross section $\sigma_T$ and
the amplitude $Q$ of the CMB quadrupole. The result is
(see \cite{Holder} for details):
\bea \label{result}
\frac{P}{Q} \, &\simeq& \, \frac{24 \pi}{25}  \bigl( \frac{3}{4 \pi} \bigr)^{1/2} \sigma_T f(t) G \mu
v \gamma(v) \\
& & \times \Omega_B \rho_c(t_0) m_p^{-1} t_0 \bigl( z(t) + 1 \bigr)^2 \bigl( z(t_i) + 1 \bigr)^{1/2} \, ,
\nonumber
\eea
where $t_i$ is the time when the wake is laid down, $\Omega_B$ is
the fraction of the total energy density which is in baryons, and
$\rho_c(t_0)$ is the critical energy density (energy density of
a spatially flat universe) at the present time $t_0$.

Note that the induced polarization increases as the wake
formation time decreases. This is because early wakes have had
time to accrete more matter and are thicker than late wakes.
The value of $P$ also increases as $z(t)$ increases. This is
because the density in a wake is larger at earlier times than
later ones. Inserting the value of the constants appearing
above and evaluating the result in units of the characteristic values 
$z(t) + 1 = z(t_i) + 1 = 10^3$ we obtain
\be \label{value}
\frac{P}{Q} \, \sim \, f(t) G \mu v \gamma(v) \Omega_B 
\bigl( \frac{z(t) + 1}{10^3} \bigr)^2 \bigl( \frac{z(t_i) + 1}{10^{1/2}} \bigr)^3
10^7 \, . 
\ee
The residual ionization fraction drops off after
recombination to a value of between $10^{-5}$ and $10^{-4}$
(see e.g. \cite{GilHolder}),
but increases again at the time of reionization to a value of order
unity. However, as can be seen from (\ref{value}), the
predicted CMB polarization has an amplitude which is suppressed
by either $f(t)$ for times $t$ between recombination and
reionization, or by the square of $(z(t) + 1) / 10^3$ for
times after reionization.

In spite of the small amplitude of the signal, the string-induced
polarization may still be detectable because of the specific
geometry of the signal in position space: a cosmic string wake
leads to a rectangular region in the sky with extra polarization
with almost uniform polarization axis and with an amplitude which
is increasing monotonically from the leading edge of the wake (where
the string is located) to the trailing edge (where the string was
at the time of wake formation), and because statistically an
equal amount of E-mode and B-mode polarization is generated.

\begin{figure}
\includegraphics[height=6cm]{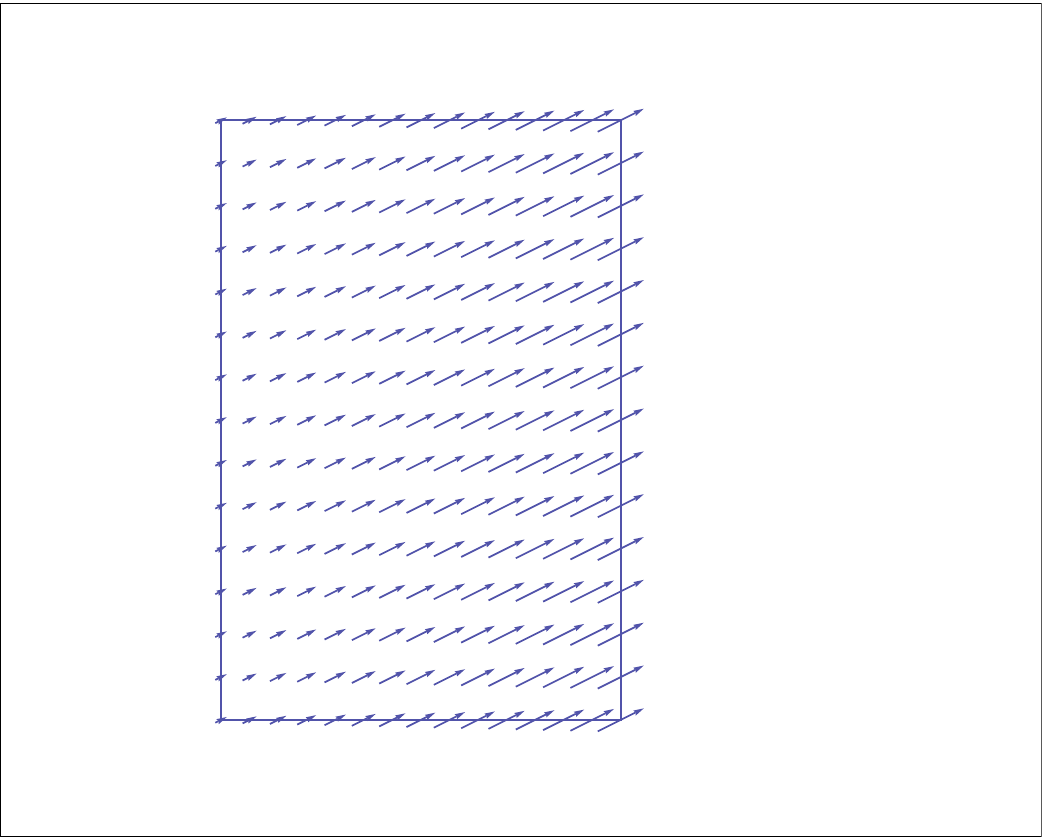}
\caption{The CMB anisotropy polarization pattern (direction and 
magnitude indicated by the arrows) produced
by a cosmic string wake. For a wake produced close to the time of
recombination, the angular size is about one degree.} 
\label{fig4}
\end{figure}

In this context, I wish to comment on the statement which is
often heard that ``B-mode polarization is the holy grail
of inflation'' \footnote{The following paragraphs are
based on \cite{holygrail}.}. What is usually meant by this statement is
that B-mode polarization is predicted in inflationary models
from the spectrum of gravitational waves which inflation
generates. Thus, it is claimed that the detection of B-mode
polarization will be a signal of the gravitational waves
which inflation predicts. However, as we have shown above,
cosmic strings predict direct B-mode polarization independent
of gravitational waves. Thus, it is not correct to interpret
a potential detection of B-mode polarization as being due
to gravitational waves.

Secondly, there are sources of gravitational waves which
generically lead to a higher amplitude of such waves than
what is obtained in the simplest single field slow-roll
inflation models. Specifically, the cosmic string loops
which are inevitably produced via the dynamics of the
scaling network of long strings will oscillate and decay
by emitting gravitational waves. Thus, a detection of gravitational
waves (via B-mode polarization or other means) is more
likely to be a signal of something different from inflation.

There is, in fact, an interesting twist to the story:
If B-mode polarization is discovered and shown to be due to
gravitational waves, and a blue spectrum (more power on
shorter wavelengths) is measured, then this would rule
out these gravitational waves as being due to inflation.
Instead, it would be the confirmation of an effect first
predicted in the context of superstring theory, namely
the spectrum induced in the string gas alternative
to inflationary cosmology \cite{Patil}.

In conclusion, B-mode polarization can be viewed as the
``holy grail'' of early universe cosmology, but not as
the ``holy grail'' of inflation.

\section{Signatures of Cosmic Strings in 21cm Surveys}
\label{21cm}

The 21cm redshift survey technique is emerging as a promising
tool to probe cosmology, in particular the high redshift universe,
the universe during the ``dark ages" (i.e. before the onset of
star formation). 

The physics is the following: after the time of recombination but
before reionization the baryonic matter in the universe is mostly
in the form of neutral hydrogen. Neutral hydrogen has a hyperfine 
transition line at a frequency corresponding for 21cm.
 If we consider the primordial CMB radiation
passing through a gas cloud, then the spectrum at the rest frame
frequency of 21cm is changed by excitation or de-excitation
of the hyperfine transition. The primordial 21cm photons can
be absorbed by the gas cloud. In turn, a hot gas cloud will emit
21cm photons. Whether the net effect is an absorption or emission
effect depends on the temperature of the gas cloud.

Any gas cloud which intersects our past light cone at some
time between recombination and the present time will yield
a 21cm signal. If the time of intersection corresponds to
a redshift $z(t)$, then the emission / absorption signal will
be seen at the redshifted wavelength
\be
\lambda(t) \, = \, (z(t) + 1) \lambda_0 \, ,
\ee
where $\lambda_0$ is 21cm. Hence, 21cm redshift surveys
provide a means for mapping the distribution of baryonic
matter as a function of redshift, including at times before
the onset of star formation.

Cosmic string wakes are nonlinear density perturbations
present at arbitrarily early times with a distinctive geometric
pattern in position space. Thus, as realized in \cite{Oscar1},
(see also \cite{21early} for early work on cosmic strings and
21cm redshift surveys), cosmic strings predict striking
signals, in particular at redshifts before reionization.

Let us begin this section with some basic equations required
to compute the 21cm signal (see \cite{Furl} for a comprehehsive
review of 21cm cosmology). Let us consider primordial CMB
photons passing through a gas cloud at redshift $z(t)$. Then
the brightness of the 21cm radiation emerging from the gas cloud
is
\be \label{basic1}
T_b(\nu) \, = \, T_S \bigl( 1 - e^{- \tau_{\nu}} \bigr) + T_{\gamma}(\nu) e^{- \tau_{\nu}} \, , 
\ee
where $T_S$ is the ``spin temperature" of the hydrogen atoms in the
gas cloud, and $T_{\gamma}$ is the temperature of the CMB
photons before entering the gas cloud. The quantity $\tau_{\nu}$ is
the optical depth at the frequency being considered. The
second term in (\ref{basic1}) describes the absorption of the primordial
CMB photon due to excitation of the hydrogen atoms, the first term
gives the contribution due to de-excitation of the hydrogen atoms.

The spin temperature introduced above determines the excitation level
of the hydrogen atoms. It is related to the gas temperature in the wake
and to the CMB temperature via a collision coefficient $x_c$:
\be \label{basic2}
T_S \, = \, \frac{ 1 + x_c}{1 + x_c T_{\gamma} / T_K} T_{\gamma} \, . 
\ee

The relative brightness temperature measured by the observer today
is
\be \label{basic3}
\delta T_b(\nu) \, = \, \frac{ T_b(\nu)-T_{\gamma}(\nu) }{1 + z} \, ,
\ee
where the denominator is due to the redshifting of the temperatures
between the time when the gas cloud intersects our past light cone
and the present time.

\begin{figure}
\includegraphics[height=6cm]{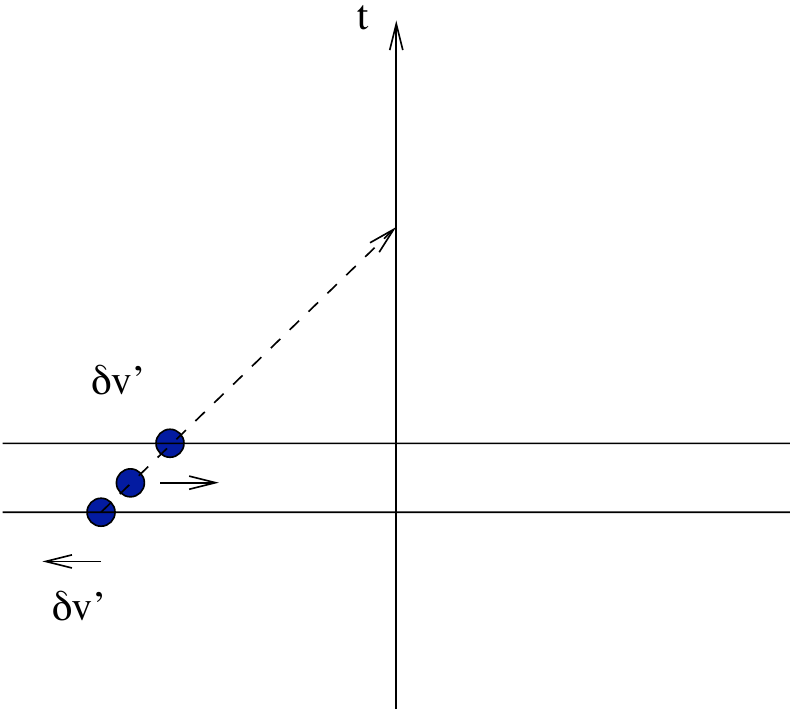}
\caption{Space-time sketch (space as horizontal axis and time in
vertical direction) showing 21cm photons emitted from
a string wake. Looking in a fixed direction in the sky, photons
from the wake will undergo different amount of cosmological redshift
and will hence arrive with a frequency dispersion.} 
\label{fig5}
\end{figure}

Let us now turn to the application to cosmic string wakes. Fig. 5
presents a sketch of 21cm photons from a string wake. As indicated,
there will be a frequency dispersion caused by the different times
the photons are emitted. Its magnitude is proportional to the width
of the wake and is given by
\be \label{freqdisp}
\frac{\delta \nu}{\nu} \, = \, 2 {\rm{sin}}(\theta) \tan({\theta}) H w \, , 
\ee
where $H$ is the Hubble expansion rate, $w$ is the wake width, and
$\theta$ is the angle between the perpendicular to the plane of the
wake and the line of sight towards us.

The optical depth for 21cm photons passing through the wake is
(in units where $c = \hbar = k_B = 1$)
\be \label{optical}
\tau_{\nu} \, = \, 
\frac{3 A_{10}}{4 \nu^2} \bigl( \frac{\nu}{T_S} \bigr) \frac{N_{HI}}{4} \phi(\nu) \, ,
\ee
where $N_{HI}$ is the column number density of hydrogen atoms, $A_{10}$ is
the spontaneous emission coefficient of the 21cm transition, and 
$\phi(\nu)$ is the line profile
\be \label{line}
\phi(\nu) \, = \, \frac{1}{\delta \nu} \,\,\, \rm{for} \,\,\, 
\nu \, \epsilon \, [\nu_{10} - \frac{\delta \nu}{2}, \nu_{10} + \frac{\delta \nu}{2}] \, .
\ee
Note that the width of the string wake (and hence the dependence on $g \mu$
cancels out between the column density and the line profile. This will
lead to the relative brightness temperature signal being independent
of the string tension.

To determine the relative brightness temperature, we also need to
know the spin temperature of the hydrogen gas inside the wake.
This is in turn determined (\ref{basic2}) from the kinetic
temperature of the wake. Assuming that the kinetic temperature
is obtained via thermalization from the kinetic energy
acquired during the Zel'dovich collapse
of the wake leads to the result \cite{Oscar1}
\be \label{waketemp}
T_K \, \simeq \,  [20~{\rm K}] (G \mu)_6^2 (v \gamma(v))^2 \frac{z_i + 1}{z + 1}  \, ,
\ee
where $(G \mu)_6$ is the value of $G \mu$ in units of $10^{-6}$, $z_i$ is
the redshift of wake formation and $z$ is the redshift at which the
wake intersects the past light cone.

Inserting this result (\ref{waketemp}) into the expression for
the 21cm brightness temperature (\ref{basic1}) yields the following
relative brightness temperature
\be
\delta T_b(\nu) \, = \, [0.07~{\rm K}] \frac{x_c}{1 + x_c} \bigl( 1 - \frac{T_{\gamma}}{T_K} \bigr) 
(1 + z)^{1/2}  
\ee
whose amplitude is of the order
\be
\delta T_b(\nu) \, \sim  200 mK    \,\,\, {\rm for} \,\,\,,\, z + 1 = 30  
\ee
(we have taken the redshift of emission from the wake to be larger
than the redshift of rionization so that we do not have to bother
about nonlinearities from the Gaussian fluctuations).

The amplitude of the cosmic string-induced 21cm signal is thus very
large. Whether the signal is in emission or absorption depends on
which of $T_K$ or $T_{\gamma}$ is larger. In the former case (low
redshifts) we have an emission signal, in the latter case it is
a signal in absorption \footnote{Note that $T_{\gamma}$ is
a decreasing function of time, whereas the wake temperature increases
for a fixed wake.}. The transition between emission and
absorption takes place when
\be
(G \mu)_6^2 \, \simeq \,  0.1 (v \gamma(v))^{-2} \, \frac{(z + 1)^2}{z_i + 1} \, .
\ee
For values of $G \mu$ which are comparable or lower than the
current upper bound, the signal is thus more likely to be in
absorption rather than in emission.

Wakes formed after the time of recombination $t_{rec}$ 
intersect the past light cone over a large area in the sky
(comparable or greater to a square degree). Each wake will
produce a signal in the form of a thin wedge in three dimensional
redshift survey space. The leading edge of the wedge (where the
cosmic string segment is located at the final time) corresponds
to a later time and hence to a frequency which has undergone
less frequency redshift. Thus, the two ``large'' dimensions
of the wedge are not exactly perpendicular to the frequency
axis. The geometry is illustrated in Fig. 6. 

\begin{figure}
\includegraphics[height=4.5cm]{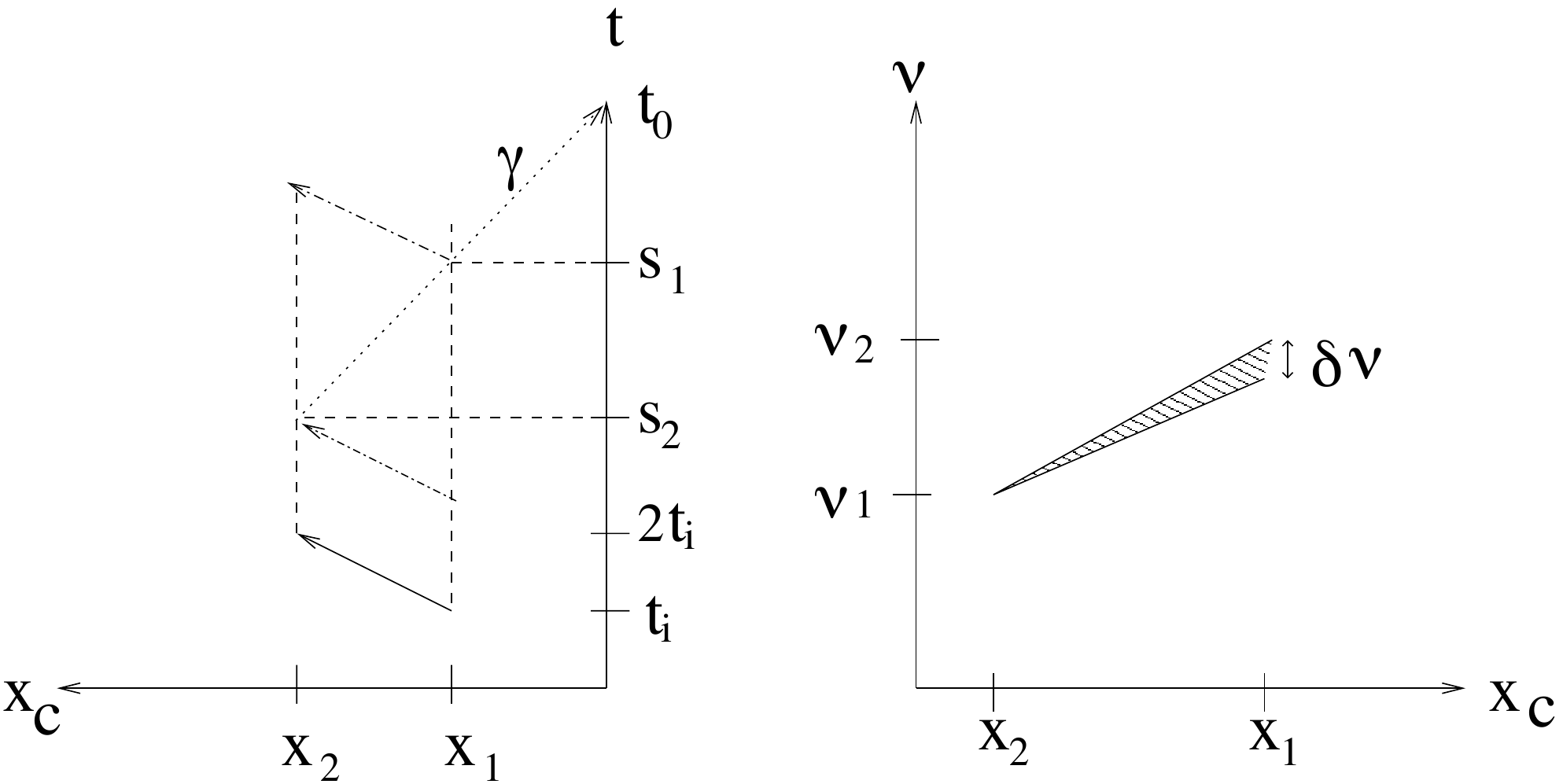}
\caption{Geometry of the 21cm signal of a cosmic string wake. On
the left is a space-time sketch (horizontal axis being
comoving spatial coordinates and vertical axis being conformal
time) of the string wake which is
laid down at time $t_i$. The string segment ``lives'' until
the time $2 t_i$. Its initial position is $x_1$, its final
position $x_2$. The string wake extends in comoving coordinates
from $x_1$ to $x_2$. Its thickness at $x_1$ vanishes and is
maximal at $x_2$. The time $t_0$ is the present time. Our
past light cone (indicated by the line labelled $\gamma$)
intersects the string wake. For the configuration shown, the
past light cone intersects the leading (thin) edge of the
wake earlier than the trailing edge. Hence, 21cm photons from
the leading edge are redshifted more than those from the trailing
edge. This gives rise to a characteristic wedge of extra
21cm absorption / emission due to the string wake in 21cm
redshift maps. This wedge is sketched on the left of the
figure, a sketch in which the horizontal axis is the same
comoving spatial coordinate as in the left sketch, and
the vertical axis is the detected 21cm frequency.} 
\label{fig6}
\end{figure}

Whereas the amplitude of the brightness temperature signal at a
fixed point inside the wedge does not depend to a first approximation
on $G \mu$, the width of the wedge does depend on it. The relative
width is given by
\bea \label{freqwidth}
\frac{\delta \nu}{\nu} \, &=& \, \frac{24 \pi}{15} G \mu v_s \gamma_s 
\bigl( z_i  + 1 \bigl)^{1/2} \bigl( z(t) + 1 \bigr)^{-1/2} \nonumber \\
& \simeq & \, 3 \times 10^{-5} (G \mu)_6 v \gamma(v)  \, , 
\eea
using $z_i + 1 = 10^3$ and $z + 1 = 30$ in the second line.
Hence, an instrument with good frequency resolution is required
to be able to measure the cosmic string wake signal at full strength.
If the frequency resolution is worse than the above value, then the
wake signal is still detectable, but with a reduced effective brightness.

The analysis presented up to this point has assumed that the initial
thermal temperature $T_g$ of the gas is negligible compared to
the temperature acquired during infall. For small values of $G \mu$
this assumption will fail. In this case the wake will not experience
shocks. The initial thermal velocities will dominate and will lead to
a wider but more diffuse wake. The width in the presence of 
initial gas temperature compared to the width $w_{T_g = 0}$ 
for $T_g = 0$  is given by \cite{Oscar2}
\be
w(t)|_{T_K < T_g} \, = \, w(t) |_{T_g = 0} \frac{T_g}{T_K} \, .
\ee
Since the hydrogen column density remains unchanged the
temperature at a fixed point in the sky will decrease since the
frequency dispersion increases. 

In Figure 7 we show the brightness temperature excess as a function
of $G \mu$ for various values of the formation redshift. For large
values of $G \mu$, the relative brightness temperature is positive.
For smaller values, it is an absorption effect. The kink point along
each curve corresponds to the lowest value of $G \mu$ for
which there is shock-heating. The (brown) almost horizontal lines on the graph
correspond to the pixel by pixel sensitivity of an experiment
such as the Square Kilometer Array (SKA), with a pixel size
optimized as a function of $G \mu$ (see \cite{Oscar2}). 
We see that the string signal is larger than the predicted noise level 
for values of $G \mu$ substantially smaller than the kink value.

\begin{figure}
\includegraphics[height=4.5cm]{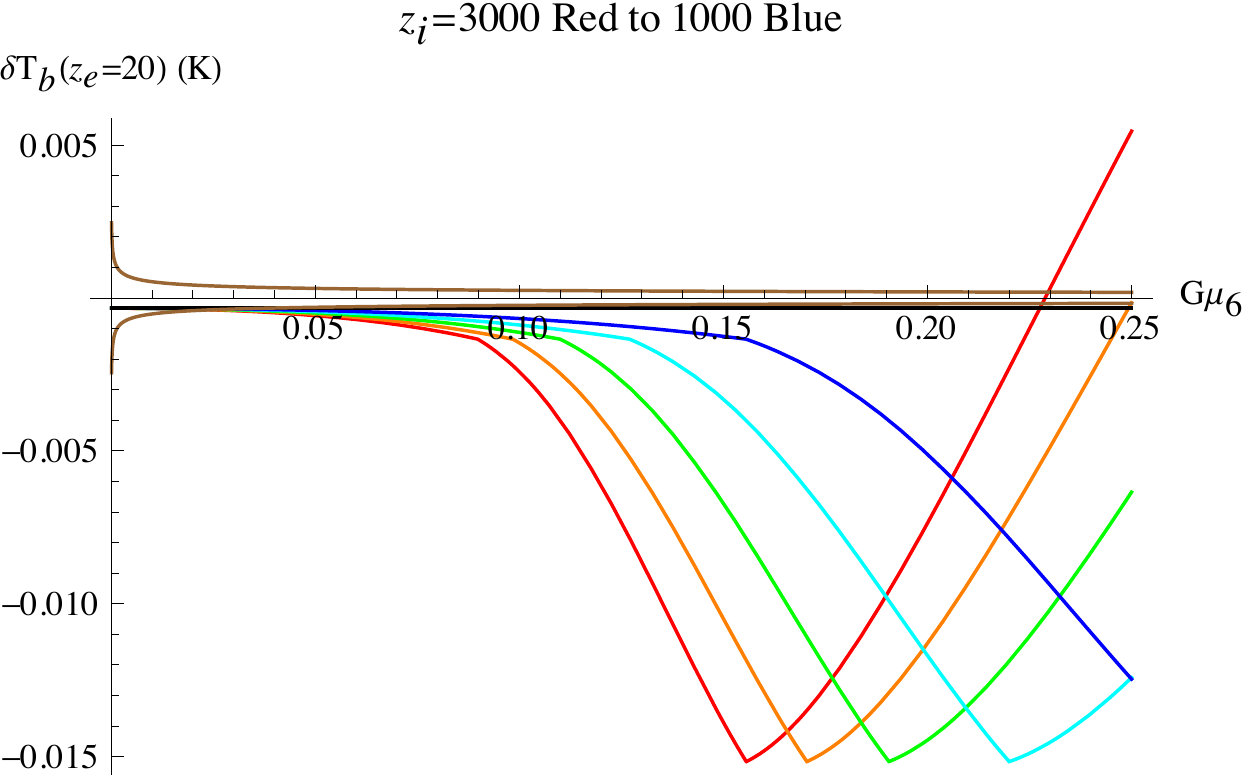}
\caption{Relative brightness temperature in degrees K induced by a cosmic
string wake for various values of the formation redshift. The curves
from left to right (in the region of low values of $G \mu$) correspond
to $z_i = 3000, 2500, 2000, 1500$ and $1000$. The horizontal axis is
the value of $G \mu$ in units of $10^{-6}$. The two brown almost horizontal
lines indicate the expected thermal noise per pixel of an experiment
such as the SKA (the $G \mu$ dependence of the horizontal lines comes
from the choice of the optimal pixel size as a function of $G \mu$ (see
\cite{Oscar2}).} 
\label{fig7}
\end{figure}

Since the signal of a string wake has a very special pattern in
position space, it is possible to search for this signal even
if the relative brightness temperature at a fixed point in the
sky is smaller than the pixel noise, in the same way that the
line discontinuities caused by string segments in CMB temperature
maps can be picked out for values of $G \mu$ where the pixel by
pixel signal is hidden in the noise. As in the case of CMB
temperature maps, it is important to use position space algorithms
to analyze the maps. For an attempt to use Minkowski functionals
to pick out string signals in 21cm redshift maps see \cite{Evan}.

Naturally, it is also possible to compute the power spectrum of
the cosmic string wake signal in 21cm redshift maps. The angular
power spectrum at a fixed value of the frequency has been
computed in \cite{Oscar3}. It is also possible to compute
the 21cm signal of a cosmic string loop. Since the overdensity
in the region which accretes around the string loop is $64$
\footnote{Assuming here that the initial translational motion
of the string loop is negligible.}
for a loop compared to $4$ for a wake (contraction occurs in
all three dimensions), the induced brightness temperature
is in fact even larger than for a string wake \cite{Pagano}. However, there
are no special patterns which allow the string loop signal to
be teased apart from background point sources. Hence, we consider
it to be more promising to search for the signals of string
wakes.

\section{Conclusions}
\label{Conclusions}

I have discussed the search for cosmic string signals in new 
observational windows. There is good motivation for this work:
detecting a cosmic string in cosmology would in itself be a
great discovery which might lead to the solution of some
outstanding problems in cosmology such as the origin of
primordial magnetic fields which are coherent on galactic
scales. Maybe more importantly, searching for cosmic strings
in the sky is a way to probe particle physics beyond the Standard
Model at energy scales which can never be reached in terrestrial
accelerators.

Cosmic strings contribute to structure formation and hence leave
imprints in the structure in the universe. However, the contribution
of strings to the total power of cosmological perturbations is already
bounded from above to be less than about $5\%$ at the present time.
The bulk of the power comes from almost Gaussian primordial 
fluctuations (here called Gaussian ``noise") such as those which 
could be generated by inflation or its alternatives. 

The first main point to take away from this talk is that cosmic strings
produce nonlinearities already at very high redshifts (whereas the
Gaussian noise does not). Hence, the signatures of cosmic strings
will be more pronounced at high redshifts. Hence, CMB anisotropy 
maps and 21cm redshift surveys are ideal windows to probe for
strings. 

The second main point is that cosmic strings produce fluctuations
which are non-Gaussian. More specifically, string wakes induce
signals with specific geometrical patterns in position space. Hence,
better limits on cosmic strings can be achieved when analyzing
the data in position space (e.g. with edge detection algorithms)
rather than via power spectra computations.

Thirdly, the 21cm redshift survey window appears to be extremely
promising. Improvements on the existing limits by several orders
of magnitude should be possible.

There are other windows to probe cosmic strings. One is via
gravitational waves. Oscillating cosmic string loops emit
gravitational radiation \cite{Tanmay}. A scaling network on strings will
produce a scale-invariant spectrum of gravitational waves
(see e.g. \cite{ABT}). Cusps on cosmic strings may produce more
distinctive signals \cite{cuspsignal}. However, it is expected that
back-reaction effects (see e.g. \cite{RHBcusp}) could greatly
reduce the gravitational wave signal of cusps, and hence limits
on $G \mu$ derived from cusp signals should be taken with
lots of grains of doubt.

Another window to probe cosmic strings is via high redshift
galaxy surveys. The relative contribution of strings to correlation
functions of nonlinear mass concentrations in the universe 
increases as the redshift grows. Hence, it is interesting to look
for signals of cosmic strings in the high redshift galaxy distribution.
Quite recently, the contribution of cosmic string loops to galaxy
formation at high redshifts has been studied \cite{Shlaer}. Corresponding
work on the effects of cosmic string wakes is ongoing at McGill.
 
\vskip0.5cm 

\centerline{\bf Acknowledgments}
 
I wish to thank Prof. Pauchy Hwang for the invitation to lecture at CosPA2012
and for his wonderful hospitality. I wish to thank all of my collaborators
on the recent cosmic string work, and in particular Rebecca Danos 
for permission to use various figures drawn for \cite{Rebecca, Holder, Oscar1},
and Oscar Hernandez for permission to use Figure 7 which is taken from
\cite{Oscar2}. This research has been supported in part by an NSERC Discovery Grant
and by funds from the Canada Research Chair program.







\end{document}